\documentstyle[12pt]{article}
\begin{document}

\large
\thispagestyle{empty}
\begin{flushright}                              FIAN/TD/96-02\\
                                                hep-th/9602048\\
                                                January 1996

\vspace{2cm}
\end{flushright}
\begin{center}
\normalsize
{\LARGE\bf  Pseudoclassical description of higher
spins in $2+1$ dimensions}

\vspace{3ex}
{\Large D. M. Gitman$^{\dagger}$}

{\large Instituto de F\'{\i}sica, Universidade de S\~ao Paulo

P.O. Box 66318, 05389-970 S\~ao Paulo, SP, Brazil}

{\Large I. V. Tyutin$^{\ddagger}$}

{\large Department of Theoretical Physics, P.~N.~Lebedev Physical
Institute,

  Leninsky prospect 53, 117924 Moscow, Russia}

\vspace{5ex}

\end{center}

\centerline{{\Large\bf Abstract}}

\normalsize

\begin{quote}
Pseudoclassical supersymmetric model to describe massive particles with
higher spins (integer and half-integer) in $2+1$ dimensions is proposed. The
quantization of the model leads to the minimal (with only one
polarization state) quantum theory. In particular, the
Bargmann-Wigner type equations for higher spins arise in course
of the canonical quantization. The cases of spin one-half and
one are considered in detail. Here  one gets Dirac particles and
Chern-Simons particles respectively. A relation with the field
theory is discussed. On the basis of the model proposed, and
using dimensional reduction considerations, a model to describe
Weyl particles with higher spins in $3+1$ dimensions is
constructed.

\end{quote}

\vfill
\noindent

$^\dagger$ E-mail address: gitman@snfma1.if.usp.br

$^\ddagger$ E-mail address: tyutin@lpi.ac.ru

\newpage


\setcounter{page}{1}

\section{Introduction}
In this paper we present  a pseudoclassical supersymmetric model to
describe massive particles with higher spins (integer and half-integer)
 in $2+1$ dimensions.
Besides a pure theoretical interest to complete the
theory of relativistic particles,  there is a
direct relation  to the $2+1$ field theory \cite{b1}, which attracts in recent
years great attention due to various reasons: e.g. nontrivial
topological properties, and especially the possibility of the
existence of particles with fractional spins and exotic statistics
(anyons), having probably applications to fractional Hall effect,
high-$T_c$ superconductivity and so on \cite{b2}. The well-known
pseudoclassical supersymmetric model for
Dirac (spin one-half) particles in $3+1$ dimensions was studied in
numerous papers \cite{BM}.
Generalizations of the model for particles with arbitrary spins in
such dimensions,
for Weyl particles, and so on, one be found, for example, in
\cite{S,GGT3,GGT2}. Attempts to
extend the pseudoclassical description to  arbitrary
odd-dimensional case had met some problems, which are connected
with the absence of an analog of $\gamma^5$-matrix in
odd-dimensions. For instance, the direct dimensional reduction of
the $3+1$ spinning particle  action (standard action) to $2+1$
dimensions does not reproduce a minimal quantum theory of the
spinning particle in $2+1$ dimensions, which has to provide only
one value of the spin projection ($1/2$ or $-1/2$). In papers
\cite{P} two modifications of the standard action were proposed
to get such a minimal theory. However, the first action \cite{P}
is in fact classically equivalent to the standard action in
$2+1$ dimensions and does not provide required quantum
properties in course of the canonical and path integral
quantization. Moreover, it is P- and T- invariant, so that an
anomaly is present. Another one does not obey gauge
supersymmetries and therefore loses the main attractive features
in such kind of models, which allow one to treat them as
prototypes of superstrings or some modes in the superstring
theory.  In \cite{GGT1} we succeeded to write a new action to
describe spin one-half in $2+1$ dimensions, which reproduces the
minimal quantum theory of this spin  after quantization. Here we
propose a model to describe all higher spins (integer and
half-integer) in $2+1$ dimensions. The action of the model is
invariant under three kinds of gauge transformations:
reparametrizations and two supertransformations. It is  P- and
T-noninvariant in full agreement with the expected properties of
the minimal theory of higher spins in $2+1$ dimensions. First,
we quantize the general model canonically, using a simple
realization in a Fock space, to demonstrate that the minimal
quantum theory of higher spins is reproduced. Then we consider
the cases of spin one-half and spin one in detail. In the first
case we present both canonical and Dirac quantizations to get
Dirac equation in $2+1$ dimensions. It turns out that  in the
case of spin one the model proposed describes Chern-Simons
particles. In particular, one can see that the equations of the
topologically massive gauge theory are reproduced in course of
the quantization. Then, in the general case, we present a
realization of the canonical quantization , which corresponds to
the Bargmann-Wigner type formulation of higher spins theory in.
A relation of the quantum mechanics
constructed with the field theory is discussed. In the end of
the paper we demonstrate that on the basis of the model
proposed, and using dimensional reduction considerations, a
model to describe Weyl particles with higher spins in $3+1$
dimensions can be constructed.

\section{The action of the model, symmetries, and Hamiltonian
formulation}

Consider pseudoclassical  action of the form
\begin{eqnarray}\label{b1}
&&S=\int_{0}^{1}\left\{-\frac{z^2}{2e}-\frac{m^2}{2}e-
\sum_{a=1}^{N}\left[sm\left( \frac{\kappa_a}
{2}+i\psi^{3}_{a}\chi_a\right)+i\psi_{an}\dot\psi^n_a\right]
\right\}d\tau
=\int_0^1Ld\tau\;, \nonumber \\
&&z^\mu=\dot x^\mu+i\sum_{a=1}^{N}\left(
\varepsilon^{\mu\nu\lambda}\psi_{a\nu}\psi_{a\lambda}\kappa_a
-\psi^\mu_a\chi_a\right);\; N=1,2,\ldots;\;\; s=\pm 1\;,
\end{eqnarray}
where the Greek (Lorentz) indices $\mu$, $\nu$, $\lambda$, run
over 0, 1, 2, whereas the Latin ones  $n, m $, run over
$0,1,2,3,$ one supposes summation over the repeated Greek and
Latin (n,m) indices (but not over the index $a$); $
\eta_{\mu\nu}=\hbox{diag}(1,-1,-1),\;
\eta_{mn}=\hbox{diag}(1,-1,-1,-1);\;\; x^\mu,\; e,\; \kappa_a$
are even and $\psi_{an}, \;\chi_a$ are odd variables;
$\varepsilon^{\mu\nu\lambda}$ is the totally antisymmetric tensor density of
Levi-Civita in $2+1$ dimensions. We
suppose that $x^\mu$ and $\psi_{a\mu}$ are $2+1$ Lorentz vectors and $e$,
$\kappa_a$, $\psi^3_a$, $\chi_a$ are scalars, so that the action (\ref{b1})
is invariant under the restricted Lorentz transformations
(but is not $P$- and $T$-invariant). It is invariant
under the reparametrizations and under other two types of
gauge transformations, one of which is a supergauge
transformation:
\begin{eqnarray}\label{b2} &&\delta x^\mu=\dot
x^\mu\xi\;,\;\; \delta e=\frac{d}{d\tau}(e\xi)\;,
\;\;\delta\psi_{an}=\dot\psi_{an}\xi\;, \;\;
\delta\chi_a=\frac{d}{d\tau}(\chi_a\xi)\;, \;\;
\delta\kappa_a=\frac{d}{d\tau}(\kappa_a\xi)\;; \\
&&\delta x^\mu=i\sum_{a=1}^{N} \psi^\mu_a \epsilon_a\;,\;\;\delta e=
i\sum_{a=1}^{N}\chi_a\epsilon_a\;,\;\;
\delta\psi^\mu_a=\frac{z^\mu}{2e}\epsilon_a\;,\;\;
\delta\psi^3_a=s\frac{m}{2}\epsilon_a,\;\;\delta\chi_a=\dot\epsilon_a\;,
\;\;\delta\kappa_a=0\;; \label{b3}\\
&&\delta x^\mu=-i\sum_{a=1}^{N}\varepsilon^{\mu\nu\lambda}\psi_{a\nu}\psi_{a\lambda}\theta_a
\;,\;\;
\delta\psi^\mu_a=\frac{1}{e}\varepsilon^{\mu\nu\lambda}z_\nu\psi_{a\lambda}\theta_a\;,
\;\;\delta\kappa_a=\dot\theta_a,\;\; \delta e=\delta\psi^3_a=\delta\chi_a=0\;,
\label{b4}
\end{eqnarray}
where $\xi(\tau),\;\theta_a(\tau)$ are even, and $\epsilon_a(\tau)$ are
odd parameters.

Going over to the Hamiltonian formulation, we introduce the
canonical momenta,
\begin{eqnarray}\label{b5}
&&\pi_\mu=\frac{\partial L}{\partial\dot x^\mu}=-\frac{1}{e}z_\mu\;,\;\;
P_e=\frac{\partial L}{\partial\dot e}=0\;, \;\;
P_{\chi_a}=\frac{\partial_rL}{\partial\dot{\chi_a}}=0\;, \nonumber\\
&&P_{\kappa_a}=\frac{\partial L}{\partial\dot{\kappa_a}}=0\;,\;\;
P_{an}=\frac{\partial_rL}{\partial\dot\psi^n_a}=-i\psi_{an}\;.
\end{eqnarray}
It follows from (\ref{b5}) that there exist primary constraints
$\Phi^{(1)}=0$,
\begin{equation}\label{b5a}
\Phi^{(1)}_1=P_e, \; \Phi^{(1)}_2=P_{\chi_a}, \;
\Phi^{(1)}_3=P_{\kappa_a},\; \Phi^{(1)}_{4}=P_{an}+i\psi_{an} \;.
\end{equation}
Constructing the total Hamiltonian $H^{(1)}$, according to the standard
procedure \cite{D,GT1}, we get $H^{(1)}=H+\lambda_A\Phi^{(1)}_A$, where
\begin{equation}\label{b6}
H=-\frac{e}{2}(\pi^2-m^2)+i\sum_{a=1}^{N}(\pi_\mu\psi^\mu_a+
sm\psi^3_a)\chi_a -i\sum_{a=1}^{N}
(\varepsilon^{\mu\nu\lambda}\pi_\mu\psi_{a\nu}\psi_{a\lambda} +
\frac{i}{2}sm)\kappa_a\;.
\end{equation}
From the consistency conditions $\dot\Phi^{(1)}=\{\Phi^{(1)},H^{(1)}\}=0$
one can find secondary constraints $\Phi^{(2)}=0$,
\begin{equation}\label{b6a}
\Phi^{(2)}_1 =
\pi^2-m^2,\;    \Phi^{(2)}_2=\pi_\mu\psi^\mu_a+sm\psi^3_a\;,\;\;
\Phi^{(2)}_3=\varepsilon^{\mu\nu\lambda}\pi_\mu\psi_{a\nu}\psi_{a\lambda}+
\frac{i}{2}s m\;,
\end{equation}
and determine $\lambda$, which correspond to
the primary constraints
$\Phi^{(1)}_{4}$. No more secondary constraints arise from the
consistency conditions and the Lagrangian multipliers, which correspond to the
primary constraints $\Phi^{(1)}_i, \;i=1,2,3$, remain undetermined. The
Hamiltonian (\ref{b6}) is proportional to the constraints.
One can go over from the initial set of constraints $\Phi^{(1)},
\Phi^{(2)}$ to the equivalent one $\Phi^{(1)},\tilde{\Phi}^{(2)}$,
where $\tilde{\Phi}{}^{(2)}=\Phi^{(2)}\left(\psi\rightarrow\tilde{\psi}=\psi+
\frac{i}{2}\Phi^{(1)}_4 \right)$.
The new set of constraints can be explicitly divided in a set of the
first-class constraints, which are ($\Phi^{(1)}_i$, $i=1,2,3$,
$\tilde{\Phi}^{(2)}$) and in a set of second-class constraints
$\Phi^{(1)}_4$.

Calculating the total angular momentum tensor $M_{\mu\nu}$, which corresponds
to the action (\ref{b1}), we get
\begin{equation}\label{b7}
M_{\mu\nu}=L_{\mu\nu}+S_{\mu\nu},\;
L_{\mu\nu}=x_{\mu}\pi_{\nu}-x_{\nu}\pi_{\mu},\;S_{\mu\nu}=i\sum_{a=1}^{N}
[\psi_{a\mu},\psi_{a\nu}]\;.
\end{equation}
The dual vector
$J^{\mu}=\frac{1}{2}\varepsilon^{\mu\nu\lambda}M_{\nu\lambda}$
together with the momentum $\pi_{\mu}$ are generators of
the $2+1$ Poincare algebra. The Pauli-Lubanski scalar $W$,
\begin{equation}\label{b8}
W=\pi_{\mu}J^{\mu}=\pi_{\mu}S^{\mu},\;\;
S^{\mu}=\frac{1}{2}\varepsilon^{\mu\nu\lambda}S_{\nu\lambda},
\end{equation}
specifies the helicity (spin) of the particles and similar
to $\pi^2$ is a Casimir operator in the case of
consideration.

To quantize the theory canonically one has to impose as much as
possible supplementary gauge conditions to the first-class
constraints.  In the case under consideration, it turns out to
be possible to impose the gauge conditions to all the first-class
constraints, excluding the constraints $\tilde{\Phi}^{(2)}_3$.
These constraints are quadratic in the grassmannian variables.
On the one hand, that circumstance makes it difficult to impose
a conjugated gauge condition, on the other hand, imposing these
constraints on state vectors does not create problems with
the Hilbert space definition since the corresponding operators
of constraints have a discrete spectrum. Thus, we shall treat
only the constraints $\tilde{\Phi}^{(2)}_3$ in  sense of the
Dirac method, fixing only the gauge freedom, which  corresponds
to two types of gauge transformations (\ref{b2}) and (\ref{b3}).
As a result  we remain only with the first-class constraints,
which are  the reduction of $\Phi^{(2)}_3$ to the rest of
constraints and gauge conditions. They can be used to specify
the physical states. All the second-class constraints form the
Dirac brackets.   The following gauge conditions $\Phi^G=0$ can
be imposed:  $\Phi^G_1=e+\zeta\pi^{-1}_0\;,\;\;
\Phi^G_2=\chi_a\;,\;\; \Phi^G_3=\kappa_a\;,\;\;
\Phi^G_4=x_0-\zeta\tau\;,\;\; \Phi^G_5=\psi^0_a\;, $
where $\zeta=-\hbox{sign}\,\pi^0\;$ (The gauge $x_0-\zeta\tau=0$ was
first proposed in \cite{GT2,GT1} as a conjugated gauge condition to
the constraint $\pi^2-m^2=0$, see there a detailed discussion of this
gauge). Using the consistency conditions
$\dot\Phi^G=0$, one can determine the Lagrangian multipliers,
which correspond to the primary constraints $\Phi^{(1)}_i,\;i=1,2,3$.
To go over to a time-independent set of constraints (to use the
standard scheme of quantization \cite{D} without  modifications
\cite{GT1}, which are necessary if the constraints depend on
time explicitly), we introduce the variable
$x^\prime_0,\;x^\prime_0=x_0-\zeta\tau$, instead of $x_0$,
without changing the rest of the variables.  That is a canonical
transformation in the space of all the variables with the
generating function $W=x_0\pi^\prime_0+\tau|\pi^\prime_0|+W_0$,
where $W_0$ is the generating function of the identity transformation
with respect to all the variables except $x^0$ and $\pi_0$.
The transformed Hamiltonian $H^{(1)\prime}$ is of the form
\begin{equation}\label{Ham}
H^{(1)\prime}=\omega+\{\Phi\},\;\;
\omega=\sqrt{{\vec\pi}^2+m^2}, \;\; {\vec\pi}^2=\pi_k\pi_k,\; k=1,2\;,
\end{equation}
where $\{\Phi\}$ are terms proportional to the constraints and
$\omega$ is the physical Hamiltonian.
Now all the constraints of the theory can be presented in the following
equivalent form:
$K=0$, $\phi=0$, $T=0$, where
\begin{eqnarray}\label{c1}
&&K=(e-\omega^{-1}\;,\; P_e\;;\;\; \chi_a\;,\; P_{\chi_a}\;;
\;\; \kappa_a\;,\;
P_{\kappa_a}\;;\;\;
x^\prime_0\;,\; |\pi_0|-\omega\;; \;\; \psi^0_a\;,\; P_{a0}),\;a=1,\ldots,N\;; \nonumber\\
&&\phi=(\pi_k\psi^k_a+sm\psi^3_a,\; P_{ad}+i\psi_{ad})\;, \;\;
k=1,2\;, \;\; d=1,2,3\;;\nonumber \\
&&T_a=\zeta\omega[\psi_{a}^2,\psi_{a}^1]+\frac{i}{2}sm\;.
\end{eqnarray}
Here $K$ and $\phi$ are  second-class constraints, whereas $T_a$
are first-class ones.  Besides, the set $K$ has the so called
special form \cite{GT1}. In this case, if we eliminate the
variables $e$, $P_e$, $\chi_a$, $P_{\chi_a}$, $\kappa_a$,
$P_{\kappa_a}$, $x^\prime_0$, $|\pi_0|$, $\psi^0_a$, and
$P_{a0}$, using the constraints $K=0$, the Dirac brackets with
respect to all the second-class constraints $(K,\phi)$ reduce to
 ones with respect to the constraints $\phi$ only. Thus, in
fact, we can only consider the variables $x^k$, $\pi_k$,
$\zeta$, $\psi^{k}_a$, $P_{ak},\;k=1,2$, and two sets of
constraints: the second-class ones $\phi$ and the first-class
ones $T$. Nonzero Dirac brackets for the independent variables
are \begin{eqnarray}\label{c2}
&&\{x^k,\pi_r\}_{D(\phi)}=\delta_{kr}\;, \;\;
\{x^k,x^r\}_{D(\phi)}=\frac{i}{\omega^2}\sum_{a=1}^{N}[\psi^k_a,\psi^r_a]\;, \;\;
\{x^k,\psi^r_a\}_{D(\phi)}=-\frac{1}{\omega^2}\psi^k_a\pi_r\;, \nonumber\\
&&\{\psi^k_a,\psi^r_b\}_{D(\phi)}=-\frac{i}{2}
(\delta_{kr}-\omega^{-2}\pi_k\pi_r)\delta_{ab}\;,
\;\;k,r=1,2\;.
\end{eqnarray}
The Dirac brackets between $J^{\mu}, \pi_{\mu}, W$,
and ${\bf \pi}^2$ have the form
\begin{eqnarray}\label{c3}
&&\{J^{\mu} ,J^{\nu}
\}_{D(\phi)}=\varepsilon^{\mu\nu\lambda}J_{\lambda},\;
\{\pi_{\mu} ,J_{\nu}
\}_{D(\phi)}=\varepsilon_{\mu\nu\lambda}\pi^{\lambda},  \nonumber \\
&&\{\pi_{\mu} ,W
\}_{D(\phi)}=\{J^{\mu} ,W\}_{D(\phi)}=\{J^{\mu} ,\pi^2\}_{D(\phi)}=0\;.
\end{eqnarray}
That means the
$2+1$ Poincare algebra with the Casimir operators $\hat{\pi}^2$
and  $\hat{W}$,
\begin{equation}\label{al}
[\hat{J}^{\mu} ,\hat{J}^{\nu}]=i\varepsilon^{\mu\nu\lambda}
\hat{J}_{\lambda},\;
[\hat{\pi}_{\mu} ,\hat{J}_{\nu}]=i\varepsilon_{\mu\nu\lambda}
\hat{\pi}^{\lambda}\;,
\end{equation}
is reproduced on the quantum level.

\section{Quantization}

\subsection{Preliminary consideration}

To verify right away that the model proposed reproduces
particles with higher spins in $2+1$
dimensions after quantization, we consider first a simple realization in a Fock
space. Then in the next subsections we present different
realization, which, however, has more close relation to the
field theory.

Let us go over to new
variables whose Dirac brackets have a simple form.  Namely,
introduce new even variables $X^k$ and odd variables
$\theta^k_a$ according to the formulas
\begin{equation}\label{c4}
X^k=x^k+\frac{i\pi_r}{m(\omega+m)}\sum_{a=1}^{N} [\psi^k_a,\psi^r_a]\;,\;\;
\theta^k_a=\psi_a^k+\frac{\pi_k(\omega-m)}{{m \vec\pi}^2}\pi_r\psi^r_a \;.
\end{equation}
Using the brackets (\ref{c2}), we get for the new variables
\begin{eqnarray}\label{c5}
&&\{X^k,\pi_r\}_{D(\phi)}=\delta_{kr}\;, \;\;
\{X^k,X^r\}_{D(\phi)}=
\{X^k,\theta^r_a\}_{D(\phi)}=\{\pi_k,\theta^r_a\}_{D(\phi)}=0\;, \nonumber\\
&&\{\theta^k_a,\theta^r_b\}_{D(\phi)}=-\frac{i}{2}\delta_{kr}\delta_{ab}\;,
\;\;k,r=1,2\;.
\end{eqnarray}
The variables $X^k,\;\pi_k,\;\zeta,\;\theta_a^k$ are independent with
respect to the second-class constraints. Thus, we remain
only with the first-class constraints (\ref{c1}), which being written in terms
of the new variables have the form
\begin{equation}\label{c6}
T_a=m\left(\zeta [\theta^2_a,\theta^1_a]+\frac{i}{2}s
\right)=0\;.
\end{equation}

The Dirac brackets (\ref{c5}) define the commutation relations
between the correspondent operators. The nonzero  commutators
(anticommutators) are \begin{equation}\label{c7}
\left[\hat{X}^k,\hat{\pi}_r\right]=i\delta_{kr}\;,
\;\;\left[\hat{\theta}^k_a,\hat{\theta}^r_b\right]_+=
\frac{1}{2}\delta_{kr}\delta_{ab}\;.
\end{equation}

We assume as usual \cite{GT2,GT1} the operator $\hat{\zeta}$ to have
the eigenvalues
$\zeta=\pm1$ by analogy with the classical theory, so that $\hat{\zeta}^2=1$,
and also we assume the equations of the second-class constraints
$\hat{\phi}=0$. Then one can
realize the algebra of all the independent  operators in a Hilbert
space ${\cal R}$,
whose elements ${\bf \Psi}\in {\cal R}$
are two-component columns dependent on ${\bf x}=(x^k),\; k=1,2$,
\begin{equation}\label{c8}
{\bf \Psi}=\left(\begin{array}{c}{\bf f}_+({\bf x})\\{\bf
f}_-({\bf x}) \end{array}\right)\;,  \;\;\hat{\zeta}=
\left(\begin{array}{cc}1&0\\0&-1 \end{array}\right)\;,\;\;
\hat{X}^k=x^k\;,\;\;\hat{\pi}_d=-i\partial_d\;,
\end{equation}
and   ${\bf f}_+({\bf x})$
and ${\bf f}_-({\bf x)}$ are ${\bf x}$
dependent vectors of a Fock space, which is constructed on the
base of the fermionic operators $(c^+_a,c_a)$  of creation and
annihilation, \begin{eqnarray}\label{c9}
&&\hat{c}_a=\hat{\theta}_a^1+i\hat{\theta}_a^2\;,\;\;\hat{c}^+_a=
\hat{\theta}_a^1-i\hat{\theta}_a^2\;, \nonumber \\
&&[\hat{c}_a,\hat{c}_b^+]_+=\delta_{ab},\;[\hat{c}_a,\hat{c}_b]_+=
[\hat{c}_a^+,\hat{c}_b^+]_+=0.
\end{eqnarray}
The operators $\hat{T}_a$ correspondent to the
first-class constraint (\ref{c6}) have the form
\begin{equation}\label{c10}
\hat{T}_a= im\hat{\zeta}\left(\hat{n}_a -\lambda \right),\;\;
\hat{n}=\hat{c}^\dagger_a\hat{c}_a\;,\;\;
\lambda =\frac{1-\hat{\zeta}s}{2}\;.
\end{equation}
These operators  specify  physical states:
\begin{equation}\label{c11}
\hat{T}_a{\bf \Psi}=0\Leftrightarrow
\hat{n}_{a}{\bf f}_{\zeta}=\delta _{-s,\zeta}
{\bf f}_{\zeta}, \;\;\zeta=\pm\;.
\end{equation}
Thus, ${\bf f}_{\zeta}$ are proportional to the vacuum vector
$|0>$ in the Fock space, $\hat{c}_{a}|0>=0$, or to the vector
$|N>= \hat{c}^+_1\ldots \hat{c}^+_N|0>$. On the other hand, the
state vectors ${\bf \Psi}$ have to obey the Schr\"odinger
equation, which defines their ``time'' dependence,
$(i\partial/\partial\tau-\hat{\omega}){\bf \Psi}=0,$
where the quantum
Hamiltonian $\hat{\omega}$ corresponds to the classical one $\omega$,
(\ref{Ham}).
Introducing the
physical time $x^0=\zeta\tau$ instead of the parameter $\tau$ \cite{GT2,GT1},
we can rewrite the Schr\"odinger equation in the following form
\begin{equation}\label{c12}
(i{\partial/\partial x^0}-\hat{\zeta}\hat{\omega}){\bf\Psi}(x)=0,\;\;
\hat{\omega}= \sqrt{\hat{\vec\pi}^2+m^2},\;\;
x=(x^0,\;{\bf x}).
\end{equation}
Together with
the eq. (\ref{c11}) that leads to the following structure of the
physical state vectors
\begin{equation}\label{c13} {\bf \Psi}
=\left(\begin{array}{c} f_+(x)|0> \\ f_-(x)|N>
\end{array}\right),  \;\;   i\frac{\partial}{\partial
x^0}f_{\pm}(x)=\pm\hat{\omega}f_{\pm}(x).
\end{equation}
We interpret   $f_+(x)|0>$ as the wave function of a particle and
$f_-^*(x)|N>$ as the wave function of an antiparticle. Both particles
and antiparticles have only one polarization state in full agreement with
the group analysis \cite{JN}.

What is the helicity (spin) of the
particles and antiparticles obtained? To answer this question let us use the
Pauli--Lubanski operator $\hat{W}$ which corresponds to the scalar
 (\ref{b8}). In the gauge selected and in the realization in
question,  it has the form
\begin{equation}\label{c14}
\hat{W}=i\hat{\zeta} m\sum_{a=1}^{N}
[\hat{\theta}_a^2,\hat{\theta}_a^1]=
\hat{\zeta} m\left(\frac{N}{2}-\hat{n}\right),\;\;\hat{n}=\sum_{a=1}^{N}
\hat{n}_a.
\end{equation}
One can easily see that the state vectors (\ref{c13}) are
eigenfunctions for the operator (\ref{c14}),
\begin{equation}\label{c15}
\hat{W}{\bf \Psi}=m\frac{sN}{2}{\bf \Psi}\;.
\end{equation}
The latter means that the spin of the particles and antiparticles
is equal to. Thus, we see that the action (\ref{b1})
describes particles with helicity (spin) $sN/2$.
It is important to compare the quantum
mechanics constructed with the field theory. To this end,
however, another realization is more convenient. We consider it
in the two next subsections.

\subsection{Spin one-half case}

Let us consider particles with spin one-half. The
corresponding model follows from the general expression (\ref{b1}) at
$N=1$. The canonical quantization considered above gives a quantum
mechanics, which completely corresponds to our ideas about the
Dirac particles in such dimensions. To get a relation with the
corresponding field theory let us consider a special realization
for initial variables $x^k$ and $\psi^k$.

It follows from the Dirac brackets (\ref{c2})  for
$N=1$ that nonzero  commutation relations have the form
\begin{eqnarray*}
&&[\hat{x}^k,\hat{\pi}_r]=i\delta_{kr}\;, \;\;
[\hat{x}^k,\hat{x}^r]=-\frac{1}{\hat{\omega}^2}[\hat{\psi}^k,\hat{\psi}^r]\;, \;\;
[\hat{x}^k,\hat{\psi}^r]=-\frac{i}{\hat{\omega}^2}
\hat{\psi}^k\hat{\pi}_r\;, \\
&&[\hat{\psi}^k,\hat{\psi}^r]_+=\frac{1}{2}
(\delta_{kr}-\hat{\omega}^{-2}\hat{\pi}_k\hat{\pi}_r)\;,
\;\;k,r=1,2\;.
\end{eqnarray*}
One can realize the algebra of all the independent operators  in a
Hilbert space ${\cal R}$
whose elements ${\bf \Psi}\in {\cal R}$
depend on ${\bf x}=(x^d)$, $d=1,2$, and have a form
\[
{\bf \Psi}({\bf x})=\left(\begin{array}{c}\Psi^{(+)}
({\bf x})\\\Psi^{(-)}({\bf x}) \end{array}\right)\;,
\]
where $\Psi^{(\pm)}({\bf x})$  are
two-component spinors, $\Psi^{(\pm)}_{\alpha}({\bf
x}),\;\alpha=1,2$. In this space
\begin{eqnarray}\label{33}
&&\hat{x}^k=x^k+\Delta \hat{x}^k,\;\; \Delta \hat{x}^k=
\frac{1}{2\hat{\omega}^2}\varepsilon^{kr}
\left(\hat{\pi}_r\Sigma^3-sm\Sigma^r\right)\;,\nonumber \\
&&\hat{\pi}_k=-i\partial_k\;,\;\; \hat{\zeta}=\left(
\begin{array}{cc}
1&0 \\
0&-1
\end{array} \right) \;,\;\;\Sigma^k=\hbox{diag}(
\sigma^k,\;\sigma^k)\;, \nonumber\\
&&\hat{\psi}^k=\frac{1}{2}\left[\Sigma^k-
\frac{\hat{\pi}_k}{\hat{\omega}^{2}}
(\hat{\pi}_r
\Sigma^r+sm\Sigma^3)\right], \;\;   k,r=1,2\;,
\end{eqnarray}
where $\sigma^k$ are Pauli matricies.
Constructing the operator $\hat{T}$ according  to the
first-class constraint (\ref{c1}) at $N=1$, we specify the
physical states,
\begin{equation}\label{d7}
\hat{T}{\bf
\Psi}=0\;,\;\; \hat{T}=\frac{i
sm\Sigma^3}{2\hat{\omega}}\hat{\zeta}\left[
\hat{\zeta}\hat{\omega}\Sigma^3+i\partial_1(i\Sigma^2)+
i\partial_2(-i \Sigma^1)-sm\right]\;.
\end{equation}
Besides, the Schr\"odinger
equation (\ref{c12}) holds for these states. The combination of
the latter equation with the condition (\ref{d7}) leads to the
Dirac equation in $2+1$ dimensions,
\begin{equation}\label{d8}
(i\partial_\mu\gamma^\mu-sm)\Psi^{(\zeta)}(x)=0\;,\;\;\zeta =\pm\;,
\end{equation}
where $\gamma^{\mu}$ are $\gamma$-matrices in $2+1$
dimensions,
\begin{eqnarray}\label{gamma}
&&\gamma^0=\sigma^3,\;\gamma^1=i\sigma^2,\;
\gamma^2=-i\sigma^1, \nonumber \\
&&[\gamma^{\mu},\gamma^{\nu}]_+=2\eta^{\mu \nu},\;\;
\gamma^{\mu}\gamma^{\nu}=\eta^{\mu\nu}-i\varepsilon^{\mu\nu\lambda}\gamma_
{\lambda}\;, \nonumber \\
&&\gamma^{+\mu}=\gamma^0\gamma^{\mu}\gamma^0,\;C\gamma^{\mu}C=-\gamma^
{T\mu},\; C=\sigma^2.
\end{eqnarray}
Calculating the operator $\hat{W}$, which corresponds to the
Pauli-Lubanski scalar  at $N=1$, we get
\begin{equation}\label{W}
\hat{W}=\frac{sm\Sigma^3\hat{\zeta}}{2\hat{\omega}}\left[sm-i\partial _1
\gamma^1-i\partial_2 \gamma^2 \right].
\end{equation}
Its action on the states, which obey the equations of motion
 (\ref{c12}), (\ref{d7}), gives the spin $s/2$ for the particles,
\begin{equation}\label{W1}
\hat{W}{\bf \Psi}(x)=m\frac{s}{2}{\bf \Psi}(x)\;.
\end{equation}
One can also verify that the operators $\hat{M}_{\mu\nu}$,
constructed according to the expression for the angular momentum
tensor  at $N=1$, act on the mass shell as Lorentz
transformations generators,
\begin{equation}\label{gen}
\hat{M}_{\mu\nu}{\bf
\Psi}(x)=\left
\{-i(x_\mu\partial_\nu-x_\nu\partial_\mu)-\frac{i}{4}
\left(\begin{array}{cc} [\gamma_{\mu},\gamma_{\nu}] &0 \\ 0 &
[\gamma_{\mu},\gamma_{\nu}] \end{array}\right)\right\}{\bf
\Psi}(x)\;.  \end{equation} It follows from the Schr\"odinger
equation that $\Psi^{(\pm)}$ can be interpreted as  positive and
negative frequency solutions to this equation.  Thus, a natural
interpretation of the components $\Psi^{(\zeta)}(x)$ is the
following:  $\Psi^{(+)}(x)$ is the wave function of a particle
with the spin $s/2$ and $\Psi^{*(-)}(x)$ is the wave function of
an antiparticle with the spin $s/2$. Such an interpretation can
be confirmed if we introduce in the model the interaction with
an external electromagnetic field, namely, if we add to the
Lagrangian of the model the following terms
\[
-g\dot{x}^{\mu}A_{\mu} + igeF_{\mu\nu}\psi^{\mu}\psi^{\nu}\;,
\]
where $g$ is the $U(1)$-charge. In this case the coupling
constants with the external field in the equations for the wave
functions  have different sign, for particles $g$ and for
antiparticles $-g$.

It is also instructive to demonstrate that the canonical
quantization considered is equivalent to the Dirac one,
where the second--class
constraints $\Phi_4^{(1)}$ define the Dirac brackets and therefore
the commutation relations,
whereas, all the first-class constraints, being applied to the
state vectors, define physical states. Thus, here we will not
impose explicitly any gauge conditions. For essential operators
and nonzero commutation relations one can obtain in the case
under consideration:
\begin{equation}\label{d1}
[\hat{x}^\mu,\hat{\pi}_\nu]=i\{x^\mu,\pi_\nu\}_{D(\Phi^{(1)}_4)}=
i\delta^\mu_\nu\;, \;\;
[\hat{\psi}^n,\hat{\psi}^m]_+=i\{\psi^n,\psi^m\}_{D(\Phi^{(1)}_4)}=
-\frac{1}{2}\eta^{nm}\;.
\end{equation}
It is possible to construct a realization of the commutation relations
(\ref{d1}) in a Hilbert space ${\cal R}$ whose elements ${\bf
\Psi}\in {\cal R}$ are four--component columns dependent on $x$,
\begin{equation}\label{d2}
{\bf \Psi}(x)=\left(\begin{array}{c}\varphi(x)\\
\Psi(x)\end{array}\right),\;\; \hat{x}^\mu=x^\mu\;,\;\;
\hat{\pi}_\mu=-i\partial_\mu\;, \;\;
\hat{\psi}^n=\frac{i}{2}\Gamma^n,
\end{equation}
where $\varphi(x)$
and $\Psi(x)$ are two-component columns, and
$\Gamma^n$, $n=0,1,2,3,$ are $\gamma$-matrices
in $3+1$ dimensions, which we select in the spinor representation
$\Gamma^0={\rm antidiag}(I,\;I)$, $\Gamma^i={\rm antidiag}
(\sigma^i,\;-\sigma^i)$, $i=1,2,3$.
According to the scheme of quantization chosen, the operators of
the first-class constraints have to be applied to the state
vectors to define the physical sector, namely, the physical states
obey the equations $\hat{\Phi}^{(2)}{\bf \Psi}(x)=0\;$, where
$\hat{\Phi}^{(2)}$ are operators, which correspond to the
constraints (\ref{b6a}). Taken into account
(\ref{d2}), one can write the equation $\hat{\Phi}^{(2)}_2{\bf
\Psi}(x)=0$ as
\begin{equation}\label{d3}
(i\partial_\mu\Gamma^\mu-sm\Gamma^3){\bf \Psi}(x)=0 \;\;
\Longleftrightarrow\left\{\begin{array}{c}
(i\partial_\mu\gamma^\mu-sm)\Psi(x)=0\;, \nonumber\\
(i\partial_\mu\gamma^{\dagger\mu}+sm)\varphi(x)=0\;.
\end{array}\right.
\end{equation}
Constructing the operator $\hat{\Phi}^{(2)}_1$ according to the
classical function $\Phi^{(2)}_1$, we discover that the equation
$\hat{\Phi}^{(2)}_1{\bf f}=0$ is not independent,
since in this case $\hat{\Phi}^{(2)}_1=
(\hat{\Phi}^{(2)}_2)^2$.
The equation $\hat{\Phi}^{(2)}_3{\bf f}(x)=0$ can be presented in the
following form
\begin{equation}\label{d4}
\left(\frac{i}{2}\varepsilon^{\mu\nu\lambda}\partial_\mu
\Gamma_\nu\Gamma_\lambda+ism\right){\bf \Psi}(x)=0 \;\;
\Longleftrightarrow \;\;
\left\{\begin{array}{c}
(i\partial_\mu\gamma^\mu-sm)\Psi(x)=0\;, \\
(i\partial_\mu\gamma^{\dagger\mu}-sm)\varphi(x)=0\;.
\end{array}\right.
\end{equation}
Combining eqs. (\ref{d3}) and (\ref{d4}), we get
$\varphi(x)\equiv0$ and $\Psi(x)$ obeys the $2+1$ Dirac equation
\begin{equation}\label{d5}
(i\partial_\mu\gamma^\mu-sm)\Psi(x)=0\;,
\end{equation}

To interpret the quantum mechanics constructed one has to take into account
the operator, which corresponds to the angular momentum tensor,
\[
\hat{M}_{\mu\nu}=-i(x_\mu\partial_\nu-x_\nu\partial_\mu)-\frac{i}{4}
\left(\begin{array}{cc}
[\gamma^\dagger_\mu,\gamma^\dagger_\nu] & 0 \\
0 & [\gamma_\mu,\gamma_\nu]
\end{array}\right)\;.
\]
Thus one can see that in fact the quantum mechanical states
are described by the two component wave function
$\Psi (x)$, which obeys the Dirac equation in $2+1$ dimensions
and is transformed under the spinor representation of the
corresponding Lorentz group.

\subsection{Bargmann--Wigner type realization}

By analogy with the canonical quantization presented above for spin one--half
case one can fulfil a quantization for arbitrary higher
spin, which leads to the Bargmann--Wigner type wave
equations \cite{BW}. Let us depart from the
Dirac brackets (\ref{c2}), which imply the following nonzero
commutation relations \begin{eqnarray}\label{36a}
&&[\hat{x}^k,\hat{\pi}_r]=i\delta_{kr}\;, \quad
[\hat{x}^k,\hat{x}^r]=-\frac{1}{\hat{\omega}^2}\sum_{a=1}^N
[\hat{\psi}^k_a,\hat{\psi}^r_a]\;, \quad
[\hat{x}^k,\hat{\psi}^r_a]=-\frac{i}{\hat{\omega}^2}
\hat{\psi}^k_a\hat{\pi}_r\;, \nonumber \\
&&[\hat{\psi}^k_a,\hat{\psi}^r_b]_+=\frac{1}{2}
(\delta_{kr}-\hat{\omega}^{-2}\hat{\pi}_k\hat{\pi}_r)\delta_{ab}\;,
\;\;k,r=1,2\;.
\end{eqnarray}
We can realize now the algebra of all the operators in a Hilbert space
${\cal R}$ whose elements ${\bf\Psi}\in{\cal R}$ depend on ${\bf
x}=(x^d), d=1,2$, and have the form
\begin{equation}\label{36A}
{\bf \Psi}({\bf x})=\left(\begin{array}{c}\Psi^{(+)}
({\bf x})\\ \Psi^{(-)}({\bf x}) \end{array}\right)\;,
\end{equation}
where each component $\Psi^{(\pm)}({\bf x})$ has $N$ spinor indices $\alpha$,
$\Psi^{(\pm)}({\bf x})=\Psi^{(\pm)}({\bf x})_{\alpha_1\ldots\alpha_N}$,
$\alpha_a=1,2$. In this space
\begin{eqnarray}\label{36b}
&&\hat{x}^k=x^k+\sum_{a=1}^N\left(\prod_{j=1}^{a-1}\otimes1\right)
\otimes\Delta\hat{x}^k\otimes\left(\prod_{j=a+1}^{N}\otimes1\right)\;,
\nonumber\\
&&\hat{\pi}_k=-i\partial_k\;,\quad\hat{\zeta}=\left(
\begin{array}{cc}1&0\\0&-1\end{array}\right)\;,\nonumber\\
&&\hat{\psi}^k_a=\left(\prod_{j=1}^{a-1}\otimes1\right)\otimes\hat{\psi}^k
\otimes\left(\prod_{j=a+1}^{n}\otimes
{1\over\hat{\omega}}(\hat{\pi}_r\Sigma^r+sm\Sigma^3)\right)\;,
\end{eqnarray}
where the operators $\Delta\hat{x}^k$ and $\hat{\psi}^k$ are
defined in (\ref{33}). The operators $\hat{T}_a$, which
correspond to the first--class constraints (\ref{c1}), have the form
\[
\hat{T}_a=\left(\prod_{j=1}^{a-1}\otimes1\right)
\otimes\hat{T}\otimes\left(\prod_{j=a+1}^{N}\otimes1\right)\;,
\]
where the operator $\hat{T}$ is defined in (\ref{d7}). They
specify the physical space $\hat{T}_a{\bf\Psi}=0$. Due to the
Sr\"odinger equation, which has still the form (\ref{c12}),
the former equations imply that both
components $\Psi^{(\pm)}$ obey the Dirac equation (\ref{d8}) for
each spinor index,
\begin{equation}\label{36c}
(i\partial_\mu\gamma^\mu-sm)_{\alpha_a\alpha^\prime_a}
\Psi^{(\pm)}_{\alpha_1\ldots\alpha^\prime_a\ldots\alpha_N}(x)=0\;,\quad
\alpha_a=1,2,\quad a=1,\ldots,N\;,
\end{equation}
and therefore obey also the Klein--Gordon equation
\begin{equation}\label{36C}
(\Box+m^2)\Psi^{\pm}(x)=0\;.
\end{equation}

The operator ${\bf\hat{W}}$, which correspond to the Pauli--Lubanski
scalar (\ref{b8}), has the form
\[
{\bf\hat{W}}=\sum_{a=1}^N\left(\prod_{j=1}^{a-1}\otimes1\right)
\otimes\hat{W}\otimes\left(\prod_{j=a+1}^{N}\otimes1\right)\;,
\]
where $\hat{W}$ is defined by eq. (\ref{W}). Its action on the
mass shell is:
\[
{\bf\hat{W}}{\bf\Psi}(x)=m{sN\over2}{\bf\Psi}(x)\;.
\]
Thus, the particles described by the states (\ref{36A}) have the
helicity $sN/2$. The operators ${\bf\hat{M}_{\mu\nu}}$, correspondent
to the angular momentum tensor
reproduce the action of the Lorentz transformation generators on
the mass shell,
\[
{\bf\hat{M}_{\mu\nu}}{\bf\Psi}(x)=\left\{-i(x_\mu\partial_\nu-
x_\nu\partial_\mu)+\sum_{a=1}^N\left(\prod_{j=1}^{a-1}\otimes1\right)
\otimes\hat{M}_{\mu\nu}\otimes\left(\prod_{j=a+1}^{N}\otimes1\right)
\right\}{\bf\Psi}(x)\;.
\]
where $\hat{M}_{\mu\nu}$ is defined by eq. (\ref{gen}).
Similar to the case of spin one--half, $\Psi^{(+)}(x)$ and
$\Psi^{(-)}(x)$ are positive and negative frequency solutions to
the wave equation, so that one can interpret $\Psi^{(+)}(x)$ as
the wave function of a particle with spin $sN/2$ and
$\Psi^{(-)*}(x)$ as that of an antiparticle with
the same spin.

The equations (\ref{36c}) are $(2+1)$ analog of the
Bargmann--Wigner equations, which describe higher spins in
$(3+1)$ dimensions \cite{BW}. In contrast with the latter case here
one does not need to impose the condition of symmetry with
respect to the spinor indices. That is connected with the
unidimensionality of the spinning space on the mass shell. The
state vectors (\ref{36A}) (Bargmann--Wigner amplitudes), which
obey the Dirac equation (\ref{36c}) for each index, are
automatically symmetric in these indices. To demonstrate that we choose
two arbitrary indices $\alpha_i$ and $\alpha_j$. Then one can
write
\[
I=(\sigma^2i\partial_\mu\gamma^\mu)_{\alpha\alpha_j}
(i\partial_\nu\gamma^\nu)_{\alpha\alpha_i}\Psi^{(\pm)}_{\alpha_i\alpha_j}=
m^2\sigma^2_{\alpha_i\alpha_j}\Psi^{(\pm)}_{\alpha_i\alpha_j}
\]
in virtue of the Dirac equation (\ref{36c}) for both indices. On
the other hand, using the properties (\ref{gamma}) of the
$\gamma$--matrices, one can write
\[
I=(\sigma^2i\partial_\mu\gamma^\mu)_{\alpha_j\alpha}
(i\partial_\nu\gamma^\nu)_{\alpha\alpha_i}
\Psi^{(\pm)}_{\alpha_i\alpha_j}=
m^2\sigma^2_{\alpha_j\alpha_i}\Psi^{(\pm)}_{\alpha_i\alpha_j}=-I\;.
\]
Thus, $\sigma^2_{\alpha_j\alpha_i}\Psi^{(\pm)}_{\alpha_i\alpha_j}=0$,
that proves the symmetry of the state vectors (\ref{36A}) in any
two spinor indices.

\section{Particles with spin one}

The canonical quantization, which was done in the Sect.III for any
$N$, reproduces the quantum mechanics of particles with spin $sN/2$ and
only one polarization state. For $N=2$ we can expect to get thus
a pseudoclassical
model for particles with spin one in $2+1$ dimensions. Let us find a
relation between such a quantum mechanics and the field
theory of massive spin one particles in such dimensions. There we have
two candidates, namely, Proca theory and topologically massive gauge theory of
Chern-Simons field \cite{b1}. Below
we are going to demonstrate that the action
(\ref{b1}) at $N=2$
leads to the theory of  Chern-Simons particles in course of quantization.
To this end let us consider first
the Dirac quantization of the theory with the action
(\ref{b1}) at $N=2$. As was already mentioned in this scheme of
quantization the second-class
constraints $\Phi_4^{(1)}$ define the Dirac brackets and therefore the
commutation relations,
whereas, the first-class constraints, being applied to the state vectors,
define physical states without imposing explicitly any gauge
conditions. For essential operators and nonzero
commutation relations one can obtain in analogy with (\ref{d1}):
\begin{eqnarray}\label{e1}
&&[\hat{x}^\mu,\hat{\pi}_\nu]=i\{x^\mu,\pi_\nu\}_{D(\Phi^{(1)}_4)}=
i\delta^\mu_\nu\;, \;\;\mu ,\nu =0,1,2\;, \nonumber \\
&&[\hat{\psi}_{an},\hat{\psi}_{bl}]_+=
i\{\psi_{an},\psi_{bl}\}_{D(\Phi^{(1)}_4)}=
-\frac{1}{2}\eta_{nl}\delta_{ab};\;\;a,b=1,2;\;\;n,l=0,1,2,3\;.
\end{eqnarray}
The commutation relations
(\ref{e1}) for $\hat{x}^{\mu}$ and $\hat{\pi}_{\nu}$ can be realized
in a Hilbert space ${\cal R}_1$ whose elements are functions dependent
on $x$, so that $\hat{x}^\mu=x^\mu \;,\;\;
\hat{\pi}_\mu=-i\partial_\mu $. The commutation relations (\ref{e1})
for $\hat{\psi}_{an}$ one can realize in a Hilbert space ${\cal R}_2$,
which is a Fock space constructed by means of four kinds of Fermi
annihilation and creation operators $\hat{b}_n,\; \hat{b}_n^{+}$,
\begin{eqnarray}\label{e2}
&&\hat{b}_n=\hat{\psi}_{1n}+i\hat{\psi}_{2n},\;
\hat{b}_n^+=\hat{\psi}_{1n}-i\hat{\psi}_{2n},  \\
&&[\hat{b}_n,\hat{b}_l^+]_{+}=-\eta_{nl},\;
[\hat{b}_n,\hat{b}_l]_{+}=[\hat{b}_n^+,\hat{b}_l^+]_{+}=0.
\nonumber
\end{eqnarray}
Due to the Fermi statistics of these operators  ${\cal
R}_2$ is a finite-dimensional space  with the basis vectors
$|0>,\;|n>,\; |nl>,\;\widetilde{|n>},\;\widetilde{|0>}$, where $|0>$
is the vacuum vector, $ \hat{b}_n |0>=0$, and
\begin{eqnarray}\label{e3}
&&|n>=\hat{b}_n^+|0>,\;|nl>=\hat{b}_n^+ \hat{b}_l^+ |0>,\;
\widetilde{|n>}=\frac{1}{6}\varepsilon ^{nlcd} \hat{b}_l^+ \hat{b}_c^+
\hat{b}_d^+|0>,
\nonumber \\
&&\widetilde{|0>}=\frac{1}{24}\varepsilon ^{nlcd}
\hat{b}_n^+ \hat{b}_l^+ \hat{b}_c^+ \hat{b}_d^+|0>,\;\;n,l,c,d=0,1,2,3\;.
\end{eqnarray}
The total Hilbert space ${\cal R}$ of the quantum mechanics, we are
constructing, is the direct product of ones ${\cal R}_1$ and ${\cal
R}_2$.  The states vectors  ${\bf f}(x)\in{\cal R}$ can be presented
in the following form
\begin{equation}\label{e4}
{\bf f}(x) =f(x)|0>+f^n(x)|n> +\frac{1}{2}f^{nl}(x)|nl>
+\tilde{f}_n(x)\widetilde{|n>}+\tilde{f}(x)\widetilde{|0>} \;.
\end{equation}
The physical  vectors of the form (\ref{e4}) have
to be annulled by the operators of the first-class constraints,
\begin{eqnarray}\label{e5}
&& (\hat{\pi}_{\mu}\hat{\psi}^\mu_a+sm\hat{\psi}^3_a){\bf
f}(x)=0, \\
&&(\varepsilon^{\mu\nu\lambda}\hat{\pi}_\mu\hat{\psi}_{a\nu}\hat{\psi}_
{a\lambda}+\frac{i}{2}sm){\bf f}(x)=0, \label{e6} \\
&&(\hat{\pi}_{\mu}\hat{\pi}^{\mu}-m^2){\bf f}(x)=0.\label{e7}
\end{eqnarray}
Combining the equations (\ref{e5}), one can get
\begin{equation}\label{e8}
\hat{\pi}^n \hat{b}_n {\bf f}(x)=0,\;\;\hat{\pi}^n \hat{b}_n^+ {\bf f}(x)=0,
\end{equation}
where $\hat{\pi}_n=(\hat{\pi}_{\mu},m),\; n=0,1,2,3$, and combining
the equations (\ref{e6}), we
get
\begin{equation}\label{e9}
(i\varepsilon^{\mu\nu\lambda}\hat{\pi}_\mu
\hat{b}^+_{\nu}\hat{b}_{\lambda}-sm) {\bf f}(x)=0.
\end{equation}
Calculating  the operators
$\hat{J}^{\mu}$ and the operator $\hat{W}$,
which correspond to  the dual angular momentum vector and to the
Pauli-Lubanski scalar  (\ref{b8}) in the realization (\ref{e2}), one
can verify that the $2+1$ Poincare algebra (\ref{al}) holds and
\begin{equation}\label{e9a}
\hat{W}=\hat{\pi}_{\mu}\hat{J}^{\mu}=i\varepsilon^{\mu\nu\lambda}\hat{\pi}_\mu
\hat{b}^+_{\nu}\hat{b}_{\lambda}\;.
\end{equation}
Thus, the equation (\ref{e9}) is well known from the group theoretical
analysis \cite{JN,BBi} condition, which specifies the helicity $s$ of  particles.
The  conditions (\ref{e8})  (for the normalized
vectors of the form (\ref{e4})) lead to the following equations
\begin{eqnarray}
&&f(x)=\tilde{f}(x)=0\;,\label{e10} \\
&& \varepsilon^{lncd}\hat{\pi}_n
f_{cd}(x)=0 \;, \label{e11} \\
&&\hat{\pi}_n f^{nl}(x)=0\;,\label{e12}
\end{eqnarray}
whereas the condition (\ref{e9}) results in
\begin{eqnarray}
&&f^n(x)=\tilde{f}_n(x)=0\;,\label{e13} \\
&&i\hat{\pi}_n\left[-\eta_{cl}\varepsilon^{nlq3}f_{qd}(x)+\eta_{dl}
\varepsilon^{nlq3}f_{qc}(x)\right]-smf_{cd}(x)=0\;.\label{e14}
\end{eqnarray}
Thus, the final form of the physical state vectors is
\begin{equation}\label{e15}
{\bf f}(x) =\frac{1}{2}f^{nl}(x)|nl>,
\end{equation}
where the  functions $f^{nl}(x)$ obey the equations
(\ref{e11}, \ref{e12}, \ref{e14}). Let us analyze consequences of
these equations in
detail. First of all, it follows from the eq. (\ref{e11}) at
$l=\mu$, that
\begin{equation}\label{e16}
f_{\mu\nu}(x)=-\frac{i}{m}(\partial_{\mu}f_{3\nu}-\partial_{\nu}f_{3\mu})\;,
\end{equation}
then the same equation at $l=3$ is obeyed identically. The relation
(\ref{e16}) means, in fact, that the theory can be formulated in
terms of the vector field ${\cal F}_{\mu}(x)=f_{3\mu}(x)$ only. One can
interpret ${\cal F}_{\mu}(x)$ as a wave function of the system in the
representation of the basis  $|3\mu>$ and in $x$-representation. The
eq. (\ref{e12}) at $l=3$ results in the transversality condition for
${\cal F}_{\mu}(x)$,
\begin{equation}\label{e17}
\partial_{\mu}{\cal F}^{\mu}(x)=0\;,
\end{equation}
whereas the same equation (\ref{e12}) at $l=\mu$ in combination
with  (\ref{e17}) provides the Klein-Gordon equation for ${\cal F}_{\mu}(x)$,
\begin{equation}\label{e18}
(\Box +m^2){\cal F}_{\mu}(x)=0,\;\;\Box=\partial_{\mu}\partial^{\mu}.
\end{equation}
Thus, the condition (\ref{e7}) for
the whole state vector ${\bf f}(x)$ holds as a consequence of eq.
(\ref{e18}). At last, we get from the equation
(\ref{e14}) at $c=3,\;d=\mu$,
\begin{equation}\label{e19}
\partial_{\lambda}\varepsilon^{\lambda\mu\nu}{\cal F}_{\nu}+sm{\cal F}^{\mu}=0\;,
\end{equation}
whereas the equations (\ref{e14}) at $c=\nu,\;d=\mu$ are  obeyed identically
as consequences of (\ref{e18}) and (\ref{e19}). The transversality
condition (\ref{e17}) is consistent with (\ref{e19}). Finally, it is
easy to discover that the Pauli-Lubanski operator in the
representation considered has the form
$(\hat{W})_{\mu}^{\nu}=\partial_{\lambda}\varepsilon^{\lambda\nu\alpha}\eta_{\alpha\mu}$,
so that eq. (\ref{e19}) is the above mentioned condition, which
specifies the helicity of particles.

One can see now that the equations (\ref{e19}) are, in fact, the field
equations  of the so called ``self-dual'' free massive field theory \cite{TPN}, with the
Lagrangian
\begin{equation}\label{e20}
{\cal
L}_{SD}=\frac{1}{2}{\cal F}^*_{\mu}{\cal
F}^{\mu}-\frac{s}{2m}\varepsilon^{\mu\nu\lambda}{\cal
F}^*_{\mu}\partial _{\nu}{\cal F}_{\lambda}\;.  \end{equation}
As was remarked in \cite{DJ} this theory is equivalent to the
topologically massive gauge theory \cite{b1} with the Chern-Simons
term. Indeed, the transversality condition (\ref{e17}) can be viewed
as a Bianchi identity, which allows introducing  gauge potentials
$A_{\mu}$, namely a transverse vector may be written (in topologically
trivial space-time) as a curl,
\begin{equation}\label{e21}
{\cal F}^{\mu}=\varepsilon^{\mu\nu\lambda}\partial
_{\nu}A_{\lambda}=\frac{1}{2}\varepsilon^{\mu\nu\lambda}F_{\nu\lambda}\;,
\end{equation}
where
$F_{\nu\lambda}=\partial_{\nu}A_{\lambda}-\partial_{\lambda}A_{\nu}$
is the field strength. Thus, ${\cal F}^{\mu}$ appears to be the dual field
strength, which is a tree-component vector in $2+1$ dimensions. Then
(\ref{e19}) implies the following equations for
$F_{\mu\nu}$
\begin{equation}\label{e22}
\partial_{\lambda}F^{\lambda\mu}+s\frac{m}{2}\varepsilon^{\mu\alpha\beta}F_{\alpha\beta}=0\;,
\end{equation}
which are the field equations of the topologically massive gauge
theory with the Lagrangian
\begin{equation}\label{e23}
{\cal L}_{CS}=-\frac{1}{4}F^*_{\mu\nu}F^{\mu\nu}+s\frac{m}{4}
\varepsilon^{\mu\nu\lambda}F^*_{\mu\nu}
A_{\lambda}\;.
\end{equation}
The theory describes particles with the mass $m$ and  spin $s=\pm 1$, having
 only one polarization state, what has been noted by several
authors \cite{b1,JN}.

One can also find a relation between the Bargmann--Wigner type
realization presented in the previous section and the
description of the spin one particle in terms of the vector field.
Let $\Psi_{\alpha\beta}(x)=\Psi^{(+)}_{\alpha\beta}(x)+
\Psi^{(-)}_{\alpha\beta}(x)$, where $\Psi^{(\pm)}(x)$ are the Bargmann--Wigner
amplitudes (\ref{36A}) for $N=2$. Then
construct a vector field ${\cal F}_\mu(x)$,
\begin{equation}\label{60} {\cal
F}_\mu(x)={1\over\sqrt2}(\sigma^2\gamma_\mu)_{\alpha\beta}
\Psi_{\alpha\beta}(x)\;.
\end{equation}
The relation between ${\cal F}_\mu(x)$ and
$\Psi_{\alpha\beta}(x)$ is one--to--one correspondence,
\[
\Psi_{\alpha\beta}(x)={1\over\sqrt2}(\gamma^\mu\sigma^2)_{\alpha\beta}
{\cal F}_\mu(x)\;.
\]
Contracting the Dirac equation (\ref{36c}) with the matrices
$\sigma^2\gamma^\mu$ and the using the symmetry property of
$\Psi_{\alpha\beta}(x)$, we verify that the equation (\ref{e19})
holds for ${\cal F}^\mu(x)$.

The Lagrangian (\ref{e20}) can be rewritten in terms of the
Bargmann--Wigner amplitude $\Psi_{\alpha\beta}(x)$,
\[
{\cal L}_{SD}={s\over2m}\overline{\Psi}_{\alpha\beta}
(i\partial_\mu\gamma^\mu-sm)_{\alpha\gamma}\Psi_{\gamma\beta},\quad
\overline{\Psi}_{\alpha\beta}=\Psi^*_{\gamma\delta}
\gamma^0_{\gamma\alpha}\gamma^0_{\delta\beta}\;.
\]
Thus we get a new formulation of the ``self--dual'' theory.

It is interesting to present for comparison a pseudoclassical model,
which reproduces  the Proca theory after quantization.  Such a model
can be derived
by means of direct dimensional reduction from the corresponding
$3+1$ dimensional model \cite{S,GGT3}. The action in $2+1$
dimensions can be written as
\begin{eqnarray}\label{e24}
&&S_{Pr}=\int_{0}^{1}\left[-\frac{z^2}{2e}-\frac{m^2}{2}e-
i\sum_{a=1}^{2}\left(
m\psi^{3}_{a}\chi_a+\psi_{an}\dot\psi^n_a\right)+
\sum_{a,b=1}^{2}f_{ab}\left(\frac{i}{2}
[\psi_{an},\psi_{b}^n]+\varepsilon_{ab}\right)
\right]d\tau \;, \nonumber \\
&&z^\mu=\dot x^\mu-i\sum_{a=1}^{2}\psi^\mu_a\chi_a\;.
\end{eqnarray}
All the notations are similar to (\ref{b1}), the new even variables
$f_{ab}$ ($f_{ab}$ is antisymmetric) are only introduced and
$\varepsilon_{ab}$ is two-dimensional Levi-Civita symbol. Both
symmetries (\ref{b2}) and (\ref{b3}) remain for the
action (\ref{e24}) (with the corresponding $z$ ), and instead
of the gauge transformations (\ref{b3}) appear new ones
\begin{equation}\label{e26}
\delta x^\mu=0\;,\;\; \delta e=0,
\;\delta\psi_{a}^n=\sum_{c=1}^2 t_{ac}\psi_{c}^n, \;
\delta\chi_a=\sum_{c=1}^2 t_{ac}\chi_c, \;
\delta f_{ab}=\dot{t}_{ab}+\sum_{c=1}^2(t_{ac}f_{cb}-t_{bc}f_{ca})\;.
\end{equation}
The classical analysis shows that the total Hamiltonian is
$H^{(1)}=H+\lambda_A\Phi^{(1)}_A$ with
\[
H=-\frac{e}{2}(\pi^2-m^2)+i\sum_{a=1}^{2}(\pi_\mu\psi^\mu_a+m\psi^3_a)
\chi_a -\sum_{a,b=1}^{2}
f_{ab}\left(\frac{i}{2}[\psi_{an},\psi_{b}^n]+\varepsilon_{ab}\right)\;,
\]
and  all the constraints coincide with ones for the action (\ref{b1})
at $N=2$, only the first-class constraints $\Phi_3^{(2)}=0$ have to be
replaced by
\begin{equation}\label{e27}
\Phi_3^{(2)}=\frac{i}{2}[\psi_{1n},\psi_2^n]+1=0\;.
\end{equation}
As a result, one can perform the Dirac quantization completely in the
same way as was done  in the Sect.V. The only difference is connected
with the different form of the first-class constraint
$\Phi_3^{(2)}$. Thus, one of the conditions, which define the physical
states, namely, the condition (\ref{e6}) has to be replaced by the
condition
\begin{equation}\label{e28}
\left(i[\hat{\psi}_{1n},\hat{\psi}_2^n]+2 \right){\bf f}(x)=0
\;\Rightarrow \left(\hat{b}^+_n \hat{b}_n +2\right){\bf f}(x)=0\;.
\end{equation}
This condition results only in the equation (\ref{e13}). Thus, in the
case under consideration, the physical state vectors have the same form
(\ref{e15}), where the functions $f^{nl}(x)$ obey only the
equations (\ref{e11}-\ref{e13}). Taking into account the consequences
of these equations, one can see that the theory can be formulated in
terms of the vector field ${\cal F}_{\mu}(x)$, which obeys only the
equations
of transversality (\ref{e17}) and the Klein-Gordon equation
(\ref{e18}), those both are equivalent to the Proca equations for the
massive vector field,
\begin{equation}\label{e29}
\partial_{\lambda}F^{\lambda\mu}+m^2{\cal F}^{\mu}=0,\;\;
F_{\mu\nu}=\partial_{\mu}{\cal F}_{\nu}(x)-\partial_{\nu}{\cal F}_{\mu}(x)\;,
\end{equation}
which implies the Proca Lagrangian
\begin{equation}\label{e30}
{\cal
L}_{Pr}=-\frac{1}{4}F_{\mu\nu}F^{\mu\nu}+\frac{m^2}{2}{\cal F}_{\mu}{\cal F}^{\mu}\;.
\end{equation}
In the Proca theory in $2+1$ dimensions two of three components
${\cal F}_{\mu}(x)$
are independent, so that two polarization states are available. In
accordance with the group theoretical analysis that means that the
Proca field corresponds to a reducible spin one representation of the
Poincare group in $2+1$ dimensions.

Comparing the action (\ref{b1}) at $N=1$ and the
action (\ref{e24}), one can believe
that the necessary reduction to only one polarization state
is achieved in the pseudoclassical action (\ref{b1}) due to the presence of
terms having a structure similar to the Chern-Simons term in the field theory
action (\ref{e23}).

In the conclusion to this section one ought to remark that we
have only  demonstrated a relation between the quantum mechanics
constructed in course of the quantization and the field theory
in cases of spin $1/2$ and $1$. The same can be done by
analogy in cases of spin $3/2,\;2,$ and $5/2$, for which the
corresponding field theory is constructed \cite{fiel}.
Unfortunately, for other higher spins the problem of the field
theory construction is still open. Its solution can be related
with an appropriate choice of the higher spins description.

 \section{Weyl particles with higher spins in 3+1
dimensions}

It is interesting that the model for spin $1/2$ in $2+1$
dimensions ($N=1$) is related by means of a dimensional reduction
procedure to the model of the Weyl particle in $3+1$ dimensions,
proposed in \cite{GGT2}. The action of the latter model has the
form
\begin{eqnarray}\label{6.1}
&&S=\int_0^1\left[-\frac{z^2}{2e}-i\psi_\mu\dot\psi^\mu\right]d\tau
\;, \nonumber \\
&&z^\mu=\dot x^\mu-\varepsilon^{\mu\nu\lambda\sigma}\kappa_\nu
\psi_\lambda\psi_\sigma -i\psi^\mu\chi-
{i s\over2}\kappa^\mu\;,
\end{eqnarray}
where $\mu=0,1,2,3$, and $\eta_{\mu\nu}={\rm diag}(1,-1,-1,-1)$.
 In the gauge $\psi^0=0$ one can see that among the four constraints $T_\mu$
 of the model only one is independent. Thus, in fact, one can use
 only one component of $\kappa^\mu$ and all others put to be zero.
 In $3+1$ dimensions this violates the explicit Lorentz
 invariance on the classical level.  However in $2+1$ dimensions it
does not. So, if we make a dimensional reduction $3+1
 \rightarrow 2+1$ in the Hamiltonian and constraints of the
 model (\ref{6.1}), putting also $\pi_3=m$, $\kappa_3=\kappa$,
 whereas $\kappa^0=\kappa^1=\kappa^2=0$, then as a result of
 such a procedure we just obtain the expression for the
 Hamiltonian of the massive spin 1/2 particle in 2+1 dimensions and
 all the constraints of the latter model. In the presence of an
 electromagnetic field one has also to put $A_3=0$,
 $\partial_3A_\mu=0$ to get the same result.

 Thus, one can think that an action, which describes the Weyl
 higher spin particles in $3+1$ dimensions, can be constructed
 in analogy with the general action (\ref{b1}) in $2+1$
 dimensions.  Namely, to describe Weyl particles with higher
 (integer and half--integer) spins (helicities) one needs to
 transform the action (\ref{6.1}) into the following one:
\begin{eqnarray}\label{6.2}
&&S=\int_0^1\left[-\frac{z^2}{2e}-i\sum_{a=1}^N\psi_{a\mu}
\dot\psi^\mu_a\right]d\tau\;, \nonumber \\
&&z^\mu=\dot x^\mu-\sum_{a=1}^N(\varepsilon^{\mu\nu\lambda\sigma}
\kappa_{a\nu}\psi_{a\lambda}\psi_{a\sigma} +i\psi^\mu_a\chi_a+
{is\over2}\kappa^\mu_a)\;.
\end{eqnarray}
The hamiltonization of the theory and its quantization can be
done quite similar to that for the actions (\ref{6.1}) and
(\ref{b1}). We describe briefly here only key
formulas and steps, using the previous notations.

The primary and secondary constraints and the Hamiltonian
(the latter is proportional to the
secondary constraints) are
\begin{eqnarray}\label{6.3}
&&\Phi^{(1)}_1=P_e\;,\; \; \Phi^{(1)}_2=P_{\chi_a}\;, \;\;
\Phi^{(1)}_3=P_{\kappa^\mu_a}\;,\;\;
\Phi^{(1)}_{4}=P_{a\mu}+i\psi_{a\mu} \;, \nonumber\\
&&\Phi^{(2)}_1=\pi^2-m^2,\; \Phi^{(2)}_2=\pi_\mu\psi^\mu_a\;,\;\;
\Phi^{(2)}_3=\varepsilon^{\mu\nu\lambda\sigma}\pi_\nu
\psi_{a\lambda}\psi_{a\sigma}+\frac{i s}{2}\pi^\mu\;, \\
&&H=-\frac{e}{2}(\pi^2-m^2)+\sum_{a=1}^{N}\left[i\pi_\mu\psi^\mu_a\chi_a
-(\varepsilon_{\mu\nu\lambda\sigma}\pi^\nu\psi^{a\lambda}
\psi^{a\sigma}+\frac{is}{2}\pi_\mu)\kappa^\mu_a\right]\;.
\end{eqnarray}
After the gauge fixing,
\[
e+\zeta\pi^{-1}_0=\chi_a=\kappa_a^\mu=x_0-\zeta\tau=\psi^0_a=0\;,
\]
and transition to the variable $x^\prime_0=x_0-\zeta\tau$, we remain with the
physical Hamiltonian
\[
H=\omega=\sqrt{{\vec\pi}^2},\quad {\vec\pi}^2=\pi_k\pi_k\;,
\]
and with the variables $x^k$, $\pi_k$, $\psi^k_{a\perp}$, $k=1,2,3$,
$\pi_k\psi^k_{a\perp}=0$, which obey the Dirac brackets
\begin{eqnarray}\label{6.7}
&&\{x^k,\pi_l\}_D=\delta_{kl}\;, \;\;
\{x^k,x^l\}_D=\frac{i}{\omega^2}\sum_{a=1}^{N}[\psi^k_{a\perp},
\psi^l_{a\perp}]\;,
\;\; \{x^k,\psi^l_{a\perp}\}_D=-\frac{1}{\omega^2}\psi^k_a\pi_l\;,
\nonumber\\
&&\{\psi^k_{a\perp},\psi^l_{b\perp}\}_D=-\frac{i}{2}
\Pi^k_l(\pi)\delta_{ab}\;, \nonumber\\
&&\Pi^k_l(\pi)=\delta_{kl}-{1\over\omega^2}\pi_k\pi_l\;.
\end{eqnarray}
The only first--class constraints, which are quadratic in grassmanian
variables, have the form
\begin{equation}\label{6.8}
\Phi^{(2)}_3={i\over2}\pi_\mu T_a\;,\quad T_a=
-{2i\zeta\over\omega}\varepsilon^{klm}\pi_k
\psi^l_{a\perp}\psi^m_{a\perp}-s\;.
\end{equation}
In course of quantization the variables turn out to be operators with
commutation relations:
\begin{eqnarray}\label{6.9}
&&[\hat{x}^k,\hat{\pi}_l]=-i\delta_{kl}\;, \;\;
[\hat{x}^k,\hat{x}^l]=\frac{1}{\hat{\omega}^2}
\sum_{a=1}^{N}[\hat{\psi}^k_{a\perp},\hat{\psi}^l_{a\perp}]\;, \;\;
[\hat{x}^k,\hat{\psi}^l_{a\perp}]=\frac{i}{\hat{\omega}^2}
\hat{\psi}^k_{a\perp}\hat{\pi}_l\;, \nonumber\\
&&[\hat{\psi}^k_{a\perp},\hat{\psi}^l_{b\perp}]_+=
\frac{1}{2}\Pi^k_l(\hat{\pi})\delta_{ab}\;.
\end{eqnarray}
In terms of the physical time $x^0$ the quantum Hamiltonian is
$\hat{H}=\hat{\zeta}\hat{\omega}$.
A realization of the Hilbert space can be constructed similar to one was made
in Sect.III. Namely,
\[
{\bf \Psi}=\left(\begin{array}{c}\Psi^{(+)}
\\ \Psi^{(-)} \end{array}\right)\;,
\]
where each component $\Psi^{(\pm)}$ have $N$ spinor indices,
$\Psi^{(\pm)}=\Psi^{(\pm)}_{\alpha_1\ldots\alpha_N}$, $\alpha_a=1,2$. The
operator $\hat{\zeta}$ acts as
\[
\hat{\zeta}{\bf \Psi}=\left(\begin{array}{c}\Psi^{(+)}
\\ -\Psi^{(-)} \end{array}\right)\;.
\]
Other operators have the form:
\begin{eqnarray}\label{6.11}
&&\hat{x}^k=x^k+\sum_{a=1}^N\left(\prod_{j=1}^{a-1}\otimes1\right)
\otimes\Delta\hat{x}^k\otimes\left(\prod_{j=a+1}^{N}\otimes1\right)\;,
\nonumber\\
&&\Delta\hat{x}^k={1\over2\hat{\omega}^2}\varepsilon^{klm}
\hat{\pi}_l\Sigma^m\;, \;\;\;\hat{\pi}_k=-i\partial_k\;,
\nonumber\\
&&\hat{\psi}^k_{a\perp}=\left(\prod_{j=1}^{a-1}\otimes1\right)\otimes
\hat{\psi}^k_\perp\otimes\left(\prod_{j=a+1}^{n}\otimes
{1\over\hat{\omega}}\hat{\pi}_l\Sigma^l\right)\;,
\;\;\;\hat{\psi}^k_\perp={1\over2}\Pi^k_l(\hat{\pi})\Sigma^l\;,
\nonumber\\
&&\hat{T}_a=\left(\prod_{j=1}^{a-1}\otimes1\right)
\otimes\hat{T}\otimes\left(\prod_{j=a+1}^{N}\otimes1\right)\;,
\nonumber\\
&&\hat{T}={1\over\hat{\omega}}\gamma^0\Sigma^l\pi_l-s\;,\quad
\gamma^0=\left(\begin{array}{cc}1&0\\0&-1\end {array}\right)\;.
\end{eqnarray}
Constructed the helicity operator ${\bf\hat{W}}$  calculated in the
realization in question we get
\begin{eqnarray*}
&&{\bf\hat{W}}=-{i\over2\hat{\pi}_0}\varepsilon_{klm}\hat{\pi}_k\hat{M}_{lm}=
\sum_{a=1}^{N}\left(\prod_{j=1}^{a-1}\otimes1\right)
\otimes\hat{W}\otimes\left(\prod_{j=a+1}^{N}\otimes1\right)\;,\\
&&\hat{W}=-{1\over2\hat{\pi}_0}\hat{\pi}_k\Sigma^k=\hat{T}+s\;.
\end{eqnarray*}
Taking into account the physical states definition
$\hat{T}_a{\bf\Psi}=0$, we get
${\bf\hat{W}}{\bf\Psi}=(sN/2){\bf\Psi}$, i.e.  the
quantum mechanics constructed describes massless particles with helicity
$sN/2$.

The realization presented  can be described in a slightly
different form. Namely, the state vector ${\bf\Psi}$ is the
Dirac multispinor:
\[
{\bf\Psi}={\bf\Psi}_{\alpha_1\ldots\alpha_N}\;,\quad\alpha_a=1,2,3,4\;.
\]
It obeys both the Schr\"odinger equation
\begin{equation}\label{6.15}
(i{\partial\over\partial x^0}-
\hat{\omega}\gamma^0)_{\alpha_a\alpha^\prime_a}
{\bf\Psi}_{\alpha_1\ldots\alpha^\prime_a\ldots\alpha_N}=0\;,
\quad a=1,\ldots,N\;,
\end{equation}
and the conditions
\begin{equation}\label{6.16}
({1\over\hat{\omega}}\gamma^0\hat{\pi}_k\Sigma^k-
s)_{\alpha_a\alpha^\prime_a}
{\bf\Psi}_{\alpha_1\ldots\alpha^\prime_a\ldots\alpha_N}=0\;,
\quad a=1,\ldots,N\;.
\end{equation}
As a consequence of these equations ${\bf\Psi}$ is symmetric in all
 the indices. One can see that the equation (\ref{6.15}) is
the Dirac equation in Foldy--Wouthuysen representation and the equation
(\ref{6.16}) reproduce the Weyl condition. If we use the Foldy--Wouthuysen
transformation
\[
{\bf\Psi}^{(D)}=\prod_{a=1}^N\otimes U^\dagger{\bf\Psi}\;,
\]
\[
U={\hat{\omega}+\gamma^k\hat{\pi}_k\over\sqrt2\hat{\omega}}\;,\quad
\gamma^k=\left(\begin{array}{cc}0&\sigma^k\\-\sigma^k&0\end{array}
\right)\;.
\]
then the equations (\ref{6.15}) and (\ref{6.16}) for the vector
${\bf\Psi}^{(D)}$ appear to be
\begin{eqnarray}\label{6.17}
&&i\partial_\mu\gamma^\mu_{\alpha_a\alpha^\prime_a}
{\bf\Psi}^{(D)}_{\alpha_1\ldots\alpha^\prime_a\ldots\alpha_N}=0\;,
\nonumber\\
&&(\gamma^5-s)_{\alpha_a\alpha^\prime_a}
{\bf\Psi}^{(D)}_{\alpha_1\ldots\alpha^\prime_a\ldots\alpha_N}=0\;,
\;\;
\gamma^5=\left(\begin{array}{cc}0&1\\1&0\end{array}\right)\;,\quad
a=1,\ldots,N.
\end{eqnarray}
Thus, we have obtained the massless Dirac equation and the Weyl
condition for each index, i.e. the Bargmann-Wigner type
description of higher massless spins in $3+1$ dimensions.

One can also verify that the model (\ref{b1}) is related to the model
(\ref{6.1}) by means of a dimensional reduction, similar to the
case of spin 1/2.
\\

{\bf Acknowledgments.}

D. M. Gitman  wish to acknowledge
Brazilian foundations  CNPq  for partial financial support and
I.V.  Tyutin thanks the European Community Commission, which is
supporting him in part under the contract INTAS-94-2317.

\end{document}